\documentclass[twocolumn,showpacs,preprintnumbers,amsmath,amssymb]{revtex4}
\usepackage{graphicx}% Include figure files
\usepackage{dcolumn}% Align table columns on decimal point
\usepackage{bm}% bold math
\usepackage{color}
\begin{document}

%\preprint{APS/123-QED}

\title{Guiding slow polar molecules with a charged wire}% Force line breaks with \\

\author{M. Strebel}
\author{S. Spieler}
\author{F. Stienkemeier}
\author{M. Mudrich}
\email{Marcel.Mudrich@physik.uni-freiburg.de}
\affiliation{Physikalisches Institut, Universit\"at Freiburg, 79104 Freiburg, Germany}

\date{\today}% It is always \today, today,

\begin{abstract}
We demonstrate experimentally the guiding of cold and slow ND$_3$ molecules along a thin charged wire over a distance of $\sim 0.34\,$m through an entire molecular beam apparatus. Trajectory simulations confirm that both linear and quadratic high-field-seeking Stark states can be efficiently guided from the beam source up to the detector. A density enhancement up to a factor 7 is reached for decelerated beams with velocities ranging down to $\sim150\,$m/s generated by the rotating nozzle technique.
\end{abstract}

\pacs{Valid PACS appear here}% PACS, the Physics and Astronomy
\maketitle

\section{\label{sec:Intro}Introduction}
Considerable experimental effort is directed towards creating dense samples of cold molecules for precision measurements~\cite{Hudson}, cold chemistry experiments~\cite{Weck:2006,Bodo:2004,Willitsch:2008,Smith:2008,Krems:2005,Krems:2009}, quantum information processing~\cite{DeMille}, degenerate quantum gases with dipolar interactions~\cite{Baranov}, etc. Established techniques for generating velocity controlled or trapped cold molecules include the deceleration using time-varying electric~\cite{Bethlem1:1999}, magnetic~\cite{VanHaecke:2007,Narevicius:2008} and optical fields~\cite{Fulton:2006} and the velocity filtering of polar molecules out of an effusive source using static or time-varying electric fields~\cite{Rangwala:2003,Junglen:2004}. Alternative routes to producing cold molecules have been demonstrated utilizing the kinematics in elastically or reactively colliding molecular beams~\cite{Elioff:2003,liu:2007}.

A more general and conceptionaly simple approach to producing slow beams of cold molecules is translating a supersonic jet to low longitudinal velocities by means of a rapidly counter-rotating nozzle. This technique was demonstrated by Gupta and Hershbach~\cite{Gupta:1999, Gupta:2001} and recently improved in our group~\cite{Strebel:2010}. Using this technique we have demonstrated the production of dense beams of various atomic and molecular species with tunable velocity ranging from thousand of m/s down to $\lesssim 100\,$m/s.

Since the molecules are not confined to any external potential this technique suffers from the drawback that beam density rapidly decay due to transverse beam expansion during beam propagation from the nozzle to the interaction region. This effect can be partly compensated by installing additional guiding elements such as electrostatic quadrupole guides~\cite{Strebel:2010}. However, such extended objects cannot easily be brought close to either the nozzle or the interaction region. Besides, electrostatic quadrupole or higher multipole guides produce confining potentials only for molecules in low-field-seeking rotational states.

As an alternative guide geometry we use a thin charged wire in this work, which is spanned through the whole molecular beam apparatus from the rotating nozzle up to the quadrupole mass spectrometer (QMS) that we use as a detector. This concept was proposed by Sekatskii~\cite{Sekatskii:1995,Sekatskii:1996} and experimentally demonstrated by Loesch~\cite{Loesch:2000}. Recently, guiding of polar molecules in surface-based electrostatic potentials has been demonstrated~\cite{Xia:2008,Deng:2011}. In our arrangement we capture the trappable molecules already in the supersonic expansion region and guide them all the way close to the detector. By comparison with classical trajectory simulations we find that ND$_3$ as well as CHF$_3$ molecules are guided in high-field-seeking states that feature both a linear and a quadratic Stark effect. An enhancement of the beam density on the beam axis due to the guiding effect of up to a factor 7 is measured for decelerated beams with velocities $v\lesssim 150\,$m/s. The guiding concept is characterized in terms of optimum geometries and in terms of the applicability to slow molecules in the ground state.

\section{Experimental setup}
\label{sec:Setup}
\begin{figure}
\begin{center}{
\includegraphics[width=0.35\textwidth]{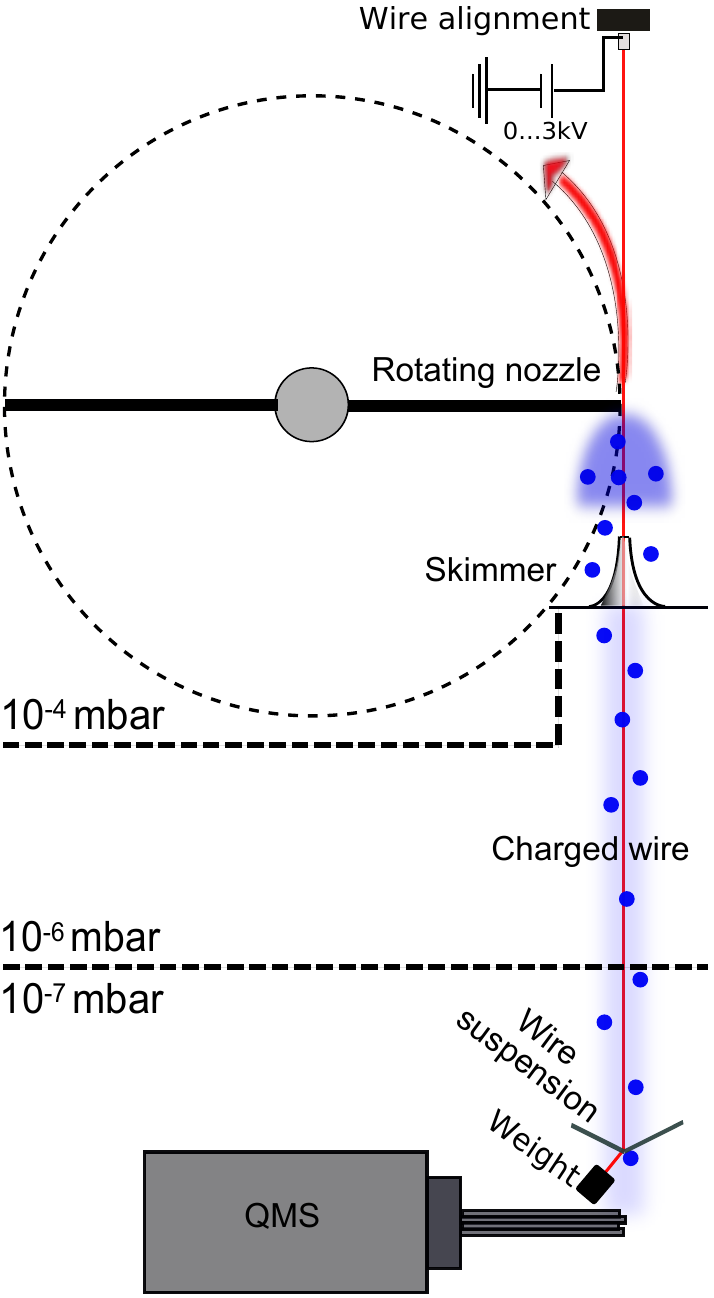}}
\caption{Schematic representation of the experimental setup used for guiding slow polar molecules along a thin charged wire.}
\label{fig:setup}
\end{center}
\end{figure}
The experimental arrangement is identical to the one reported earlier~\cite{Strebel:2010} with the difference that the quadrupole guide placed between skimmer and QMS detector is replaced by a thin wire, as schematically shown in Fig. \ref{fig:setup}. We use a gold-plated beryllium-copper (BeCu) wire with a radius $R_w=25\,\mu$m. The total length of the wire from the jet expansion region up to the suspension in front of the detector amounts to $34\,$cm. The wire is attached at one end to a $x-y$-translation stage placed inside the source chamber such that the wire can be aligned transversally with respect to the beam axis to pass close by the nozzle ($100\,\mu$m in diameter) and concentrically through the skimmer that has an aperture of 1\,mm in diameter. Further downstream the wire passes through an intermediate chamber that serves as a differential pumping section and is suspended inside the detector chamber by a perpendicularly spanned second isolated copper (Cu) wire $15\,$mm in front of the crossed beam ionizer of the QMS. The BeCu wire is held under tension by a weight of  $20$\,g such that the wire curvature at the point of suspension can be assumed to follow the radius of the Cu wire ($50\,\mu$m). Owing to the resulting kink in the electric field a large fraction of the guided molecules are output coupled and enter the detection region of the ionizer of the QMS.

Alternatively, we have installed a home-built detection unit consisting of a heated filament radially displaced from the wire at a distance of $3\,$mm and a channeltron detector on the opposite side at a distance of $35\,$mm from the wire. In this way, the molecules can be ionized directly in the guiding potential of the wire. Both the filament and the channeltron were biased to constant voltages with respect to the wire in order to ensure that the detection efficiency is independent of the wire voltage. The measured signals followed the same trends as the signals recorded using the QMS. However, a high background signal level from ionized background gas and fluctuating signal offsets caused the results to be less indicative than those measured with the QMS.

The guide wire is electrically connected to a high voltage feedthrough inside the source chamber. The maximum applicable voltage is limited to $U\approx2.2\,$kV with respect to the surrounding vacuum chambers by sparkover from the wire to either the skimmer or to the bored titanium ferrule that forms the nozzle at the tip of the rotor arm. Sparking to the skimmer can be prevented by setting the skimmer to high voltage. However, the guiding efficiency was found to diminish considerably in this case. Therefore all measurements presented here are performed with the skimmer set to ground potential and with a minimum distance $\Delta x$ between the center of the nozzle orifice and the wire center $\Delta x\approx 350\,\mu$m to avoid sparking. This, in turn, means that the nozzle is displaced away from the beam axis by the same distance since the wire is coaxially with the beam axis. As a result, the peak density of the transmitted beam without guide voltage is reduced by about a factor 4 as compared to the situation when no wire is installed and when the nozzle position is optimized ($\Delta x=0$).

\begin{figure}
\begin{center}{
\includegraphics[width=0.53\textwidth]{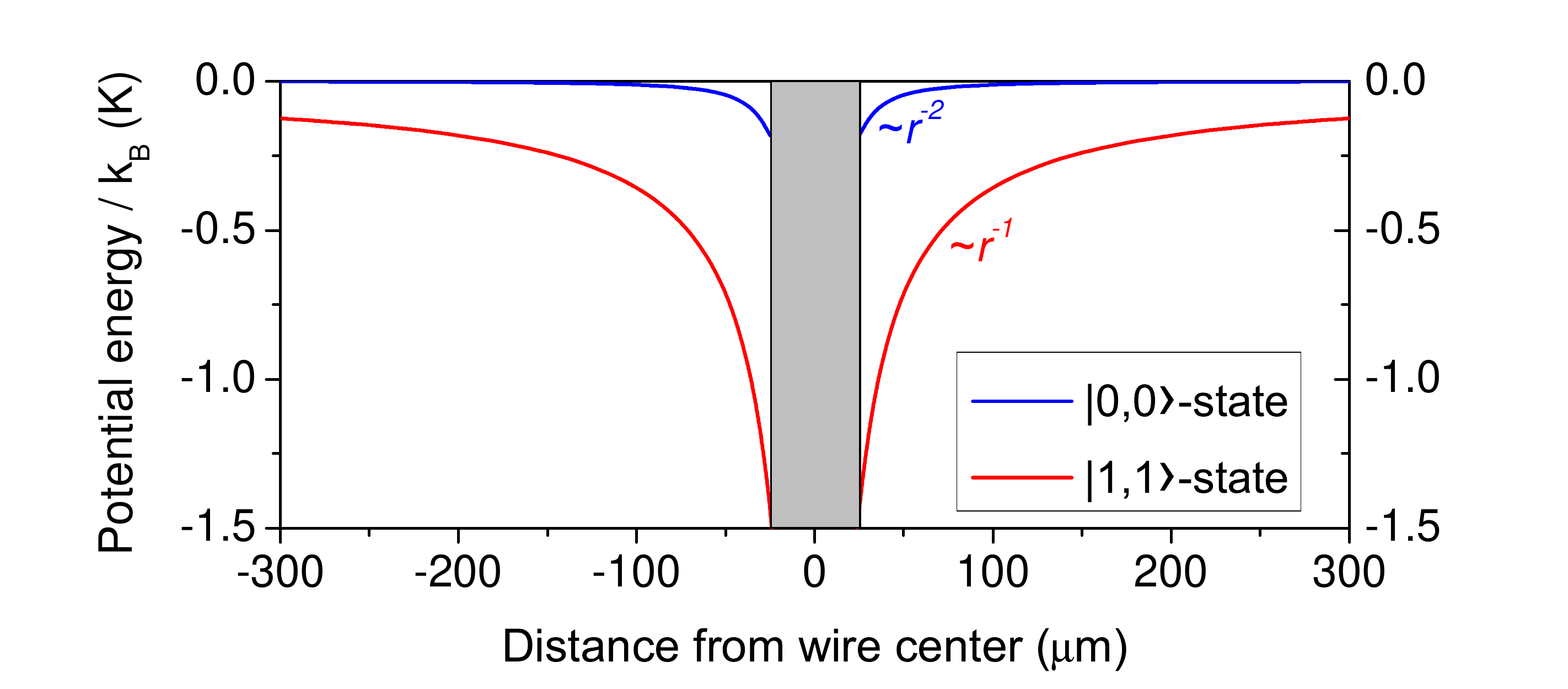}}
\caption{Transverse guide potential of ND$_3$ molecules in rotational states $|J,KM\rangle=|0,0\rangle$ (quadratic Stark effect) and $|1,1\rangle$ (linear Stark effect). A wire diameter of $50\,\mu$m and an applied voltage $U=1.5\,$kV is assumed.}
\label{fig:potential}
\end{center}
\end{figure}
In the present study we use ND$_3$ as test molecules to demonstrate the potential of the charged wire setup for guiding slow polar molecules. Since the energy difference between the vibronic ground state $J=0$ and the lowest rotationally excited state $J=1$ is about $8.3\,$cm$^{-1}$ mostly these two states are thermally populated at the estimated low rotational temperatures in the jet of $T\lesssim 10\,$K. ND$_3$ is particularly well suited for guiding and deceleration experiments using electric fields due to the strong linear Stark effect of both the high-field-seeking state $|J,KM\rangle=|1,1\rangle$ and the low-field-seeking state $(1,1,-1)$ correlating to $J=1$ rotational level which arises from the small inversion splitting in the absence of electric fields~\cite{Bethlem:2000}. The rovibronic ground state $|0,0\rangle$ is high-field-seeking and features quadratic Stark effect. When placing a ND$_3$ molecule in the electric field of a cylindrical capacitor at a distance $r$ from the center,
\begin{equation}
\label{eq:Efield}
E(r)=\frac{U}{\ln(R_0/R_w)}\frac{1}{r},
\end{equation}
created by the charged wire of radius $R_w$ inside a vacuum apparatus that is assumed to have cylindrical symmetry with inner radius $R_0\sim20\,$mm we obtain a transverse trapping potential $V$ that scales as $V(r)\propto r^{-1}$ for the linear Stark state $|1,1\rangle$ and as $V(r)\propto r^{-2}$ for the ground state $|0,0\rangle$, see Fig.~\ref{fig:potential}. The shaded area in the center of the figure indicates the range excluded by the wire. While the potential $V$ is attractive for both states in the shown range of distances $V$ is much deeper ($\sim 1.5\,$K$\times$k$_B$) and more extended for the $|1,1\rangle$-state as compared to the ground state $|0,1\rangle$. Thus, more efficient guiding of molecules in the state $|1,1\rangle$ may be expected. Moreover, molecules in $|1,1\rangle$ follow stable elliptical trajectories in analogy to the Kepler problem, whereas the $r^{-2}$-potential experienced by the ground state does not sustain stable trajectories. In this case the trajectories either spiral down to hit the wire after a few revolutions or quickly escape away from the wire, as discussed in detail by Denschlag \textit{et al.}~\cite{Denschlag:1998}. Therefore a guiding effect may be expected only for a certain range of velocities in which the molecules are transiently bound to the wire and reach the detector before crashing into the wire or being expelled away from it. This situation is discussed in more detail at the end of the following section. Note that in the experiment performed by Loesch~\cite{Loesch:2000} using alkali-halide molecules even the rotational ground states were subjected to a $r^{-1}$-potential due to ``brute force'' orientation as a result of much larger dipole moments and higher fields.

\section{Characterization of wire-guided beams}
\label{sec:theory}
\begin{figure}
\begin{center}{
\includegraphics[width=0.45\textwidth]{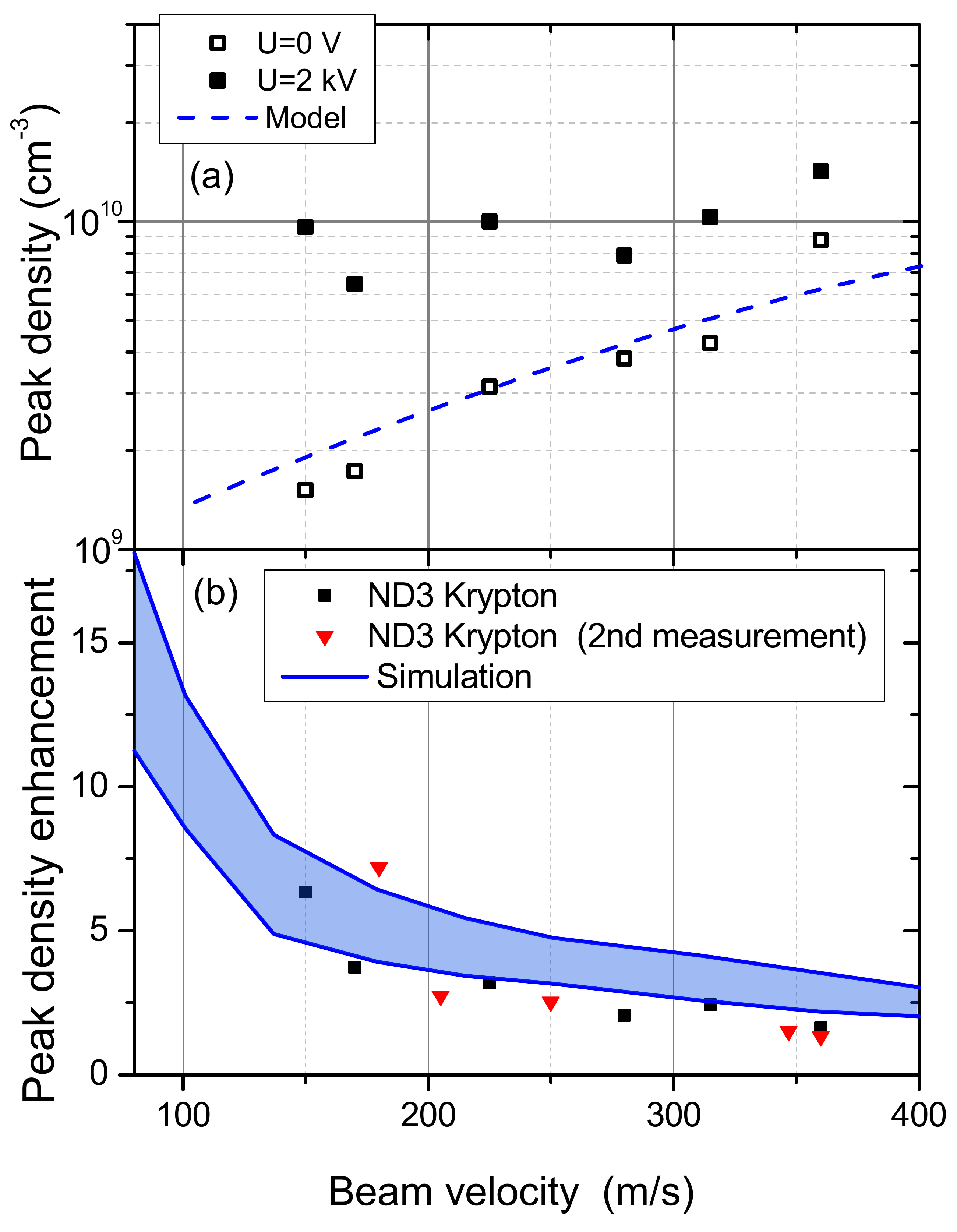}}
\caption{(a) Absolute peak densities of beams of decelerated ND$_3$ molecules detected behind the wire guide for wire voltage on (2\,kV) and off (0\,V). The dashed line represents a model of the free jet expansion based on Gaussian transverse velocity distributions. (b) Relative enhancement of the peak density due to guiding by the charged wire. The solid lines show the result of trajectory simulations (see text).}
\label{fig:velocity}
\end{center}
\end{figure}
The guiding efficiency characterized in this section is determined by comparing the detected beam density when switching on the wire voltage $U$ against the density of the unguided beam when the wire potential is set to ground ($U=0$). Such a measurement using ND$_3$ ($15\,\%$) seeded in krypton ($85\,\%$) for variable beam velocities is depicted in Fig.~\ref{fig:velocity} as open ($U=0$) and as filled ($U=2\,$kV) symbols. The drop of the absolute density of molecules (Fig.~\ref{fig:velocity} (a)) with decreasing beam velocity is due to the transverse and longitudinal dispersion of the beam. This behavior is well reproduced by simple considerations based on the expansion of a bunch of molecules that has Gaussian velocity distributions in longitudinal and transverse directions (dotted line)~\cite{Strebel:2010}. Clearly, this drop can be mostly compensated by the wire guide when applying high voltage $U$ to the wire (filled symbols). Note that guiding is particularly efficient at low velocities in proportion to the unguided beam. The relative density increase due to guiding we call enhancement, which is illustrated in Fig.~\ref{fig:velocity} (b). Thus, at beam velocities around $150\,$m/s we measure an increased beam density by up to a factor 7. A second measurement was done after the wire was replaced and newly aligned with respect to the nozzle, skimmer, and QMS detector. The two measurements are in good agreement at high beam velocities, whereas at low velocities there are slight deviations probably due to a slightly different distance $\Delta x$ between the nozzle and the wire.

The solid lines depict the result of classical trajectory simulations which account for the thermal population of rotational levels $J=0$ and $J=1$ at $T=7\,$K as well as for nuclear spin statistics~\cite{Townes:1955}. Initial values for the spatial and velocity coordinates are determined by a Monte-Carlo method according to Gaussian velocity distributions and a longitudinal spatial slit opening function of the arrangement determined from fits to the experimental time of flight measurements of the unguided beam at various beam velocities~\cite{Strebel:2010}. One data point reflects the average of 5000 trajectories bound to the wire potential for each of the 10 rotational states correlating to $J=0$ and $J=1$. The two solid lines result from the same simulation and merely reflect the uncertainty in the effective cross section of the ionizer of the QMS. The upper and lower lines correspond to estimated circular cross sections with radius $R_D=1.5\,$mm and $R_D=2\,$mm, respectively. Irrespective of this uncertainty, the simulation tends to overestimate the guiding effect. Possible detrimental effects in the experiment that are not accounted for in the simulation include modulations of the guide potential along the beam axis due to drastically changing outer radius $R_0$ of the assumed cylindrical capacitor configuration in particular when passing through the skimmer as well as the unknown efficiency of the output coupling process at the wire bend in front of the detector. According to the simulations, the enhancement of the guiding effect will rise steeply to exceed a factor of 10 as the speed of the molecules is further reduced below $\sim 150\,$m/s. Unfortunately we did not reach this velocity range with the present setup due to the limited maximum rotor frequency ($\lesssim 300\,$Hz)~\cite{Strebel:2010}.

\begin{figure}
\begin{center}{
\includegraphics[width=0.45\textwidth]{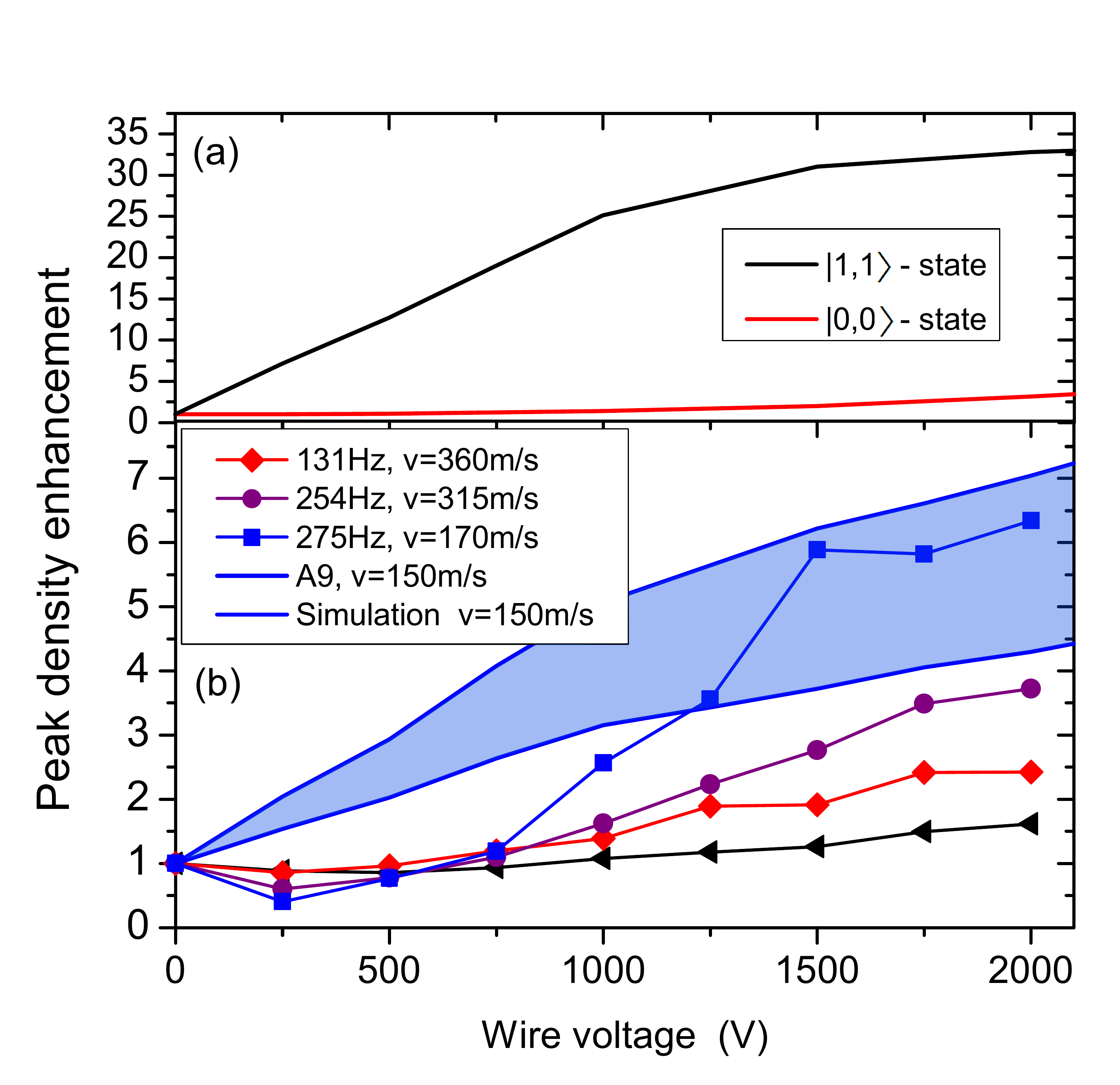}}
\caption{ (a) Simulated relative enhancement of the peak density of ND$_3$ molecules in the high- field-seeking states $|J,KM\rangle=|0,0\rangle$ and $|1,1\rangle$ due to guiding by the charged wire as a function of wire voltage for a beam velocity $v_0=150\,$m/s. (b) Enhancement of the measured ND$_3$ peak density for various beam velocities (symbols). The calculations show the average of trajectory simulations for the low energy Stark states correlating to $J=0$ and $J=1$ and for two effective detector cross sections with radius $R_D=1.5\,$mm and $R_D=2\,$mm (upper and lower lines, respectively).}
\label{fig:voltage}
\end{center}
\end{figure}
The measured enhancement of the molecule density due to guiding as a function of the wire voltage $U$ for different beam velocities $v_0$ is depicted as symbols in Fig.~\ref{fig:voltage} (b). As $U$ increases, the number of guided molecules grows monotonically, leading to a relative enhancement factor of up to 7 at $v_0=150\,$m/s. At high voltages the enhancement slightly saturates. This is due to the dominating contribution of the linear Stark state $|1,1\rangle$ which saturates at voltages $U\gtrsim 1\,$kV, shown in Fig.~\ref{fig:voltage} (a). In this voltage range nearly all molecules in the guidable $|1,1\rangle$-state are actually captured by the confining wire potential.

%Saturation in the simulated curves means the number of guided molecules remains constant, even if the potential is increased.
The quadratic Stark state $|0,0\rangle$ contributes much less in spite of its larger relative population as a consequence of the much weaker guide potential (see Fig.~\ref{fig:potential}). The solid lines in Fig.~\ref{fig:voltage} (b) depict the results of trajectory simulations for $v_0=150\,$m/s when averaging over all Stark states correlating to $J=0$ and $J=1$. Again, $R_D=1.5\,$mm and $R_D=2\,$mm are assumed (upper and lower lines, respectively). Although the trend of a monotonically increasing guiding efficiency as a function of $U$ is well reproduced, saturation sets in earlier in the simulation than in the experiment. This is presumably due to imperfections in the experimental setup as mentioned above.

\begin{figure}
\begin{center}{
\includegraphics[width=0.45\textwidth]{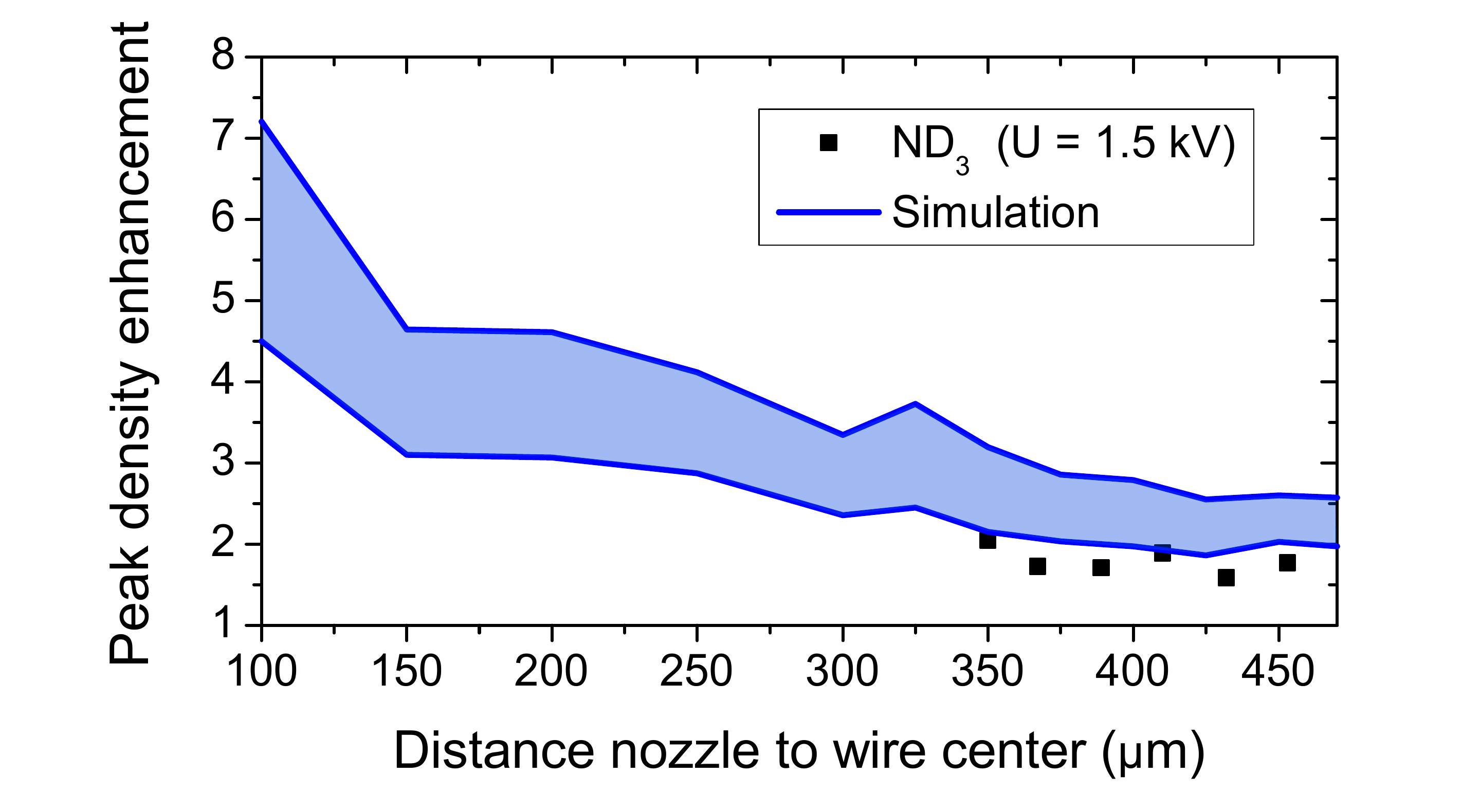}}
\caption{Measured (symbols) and simulated (lines) enhancement of the ND$_3$ peak density due to guiding by the charged wire as a function of the distance $\Delta x$ between the center of the nozzle orifice and the center of the wire. The beam velocity and wire voltage are fixed to $v_0=310\,$m/s and $U=1.5\,$kV, respectively.}
\label{fig:distance}
\end{center}
\end{figure}
The dependence of the enhancement factor on the distance $\Delta x$ between the center of the charged wire and the nozzle orifice is shown in Fig.~\ref{fig:distance}. In this measurement, the wire voltage is set to a moderate value $U=1.5\,$kV to avoid sparking, the beam velocity is held constant at $v_0=310\,$m/s, and the distance between the wire and nozzle is varied from $500\,\mu$m down to the minimum distance of about $350\,\mu$m. This is achieved by shifting the whole baseplate that supports the rotating nozzle setup with respect to the position of the charged wire which is kept fixed. In this way, the geometry of the guiding field and of the detector is maintained unchanged during the measurement. Note that the nozzle to wire distance is obtained by viewing the nozzle position through a telescope along the wire axis. Therefore the value of $\Delta x$ must be regarded as an estimate with an uncertainty of about $50\,\mu$m.

Clearly, as the nozzle is brought as close as possible to the charged wire the guiding efficiency slightly increases up to a factor of about 2. This is due to the increased depth of the guide potential close to the wire surface such that a larger fraction of molecules with non-vanishing transverse velocity components still remain bound to the wire potential. At the same time, the absolute signal of the unguided beam is increased by a factor of $1.5$ since the nozzle is moved towards the beam axis, as mentioned above. The trajectory simulations for $R_D=1.5\,$mm and $R_D=2\,$mm (solid lines) reproduce this trend, but slightly overestimate the enhancement. By further decreasing the distance, the enhancement could be increased by an additional factor of about 2 according to the simulation.

\begin{figure}
\begin{center}{
\includegraphics[width=0.45\textwidth]{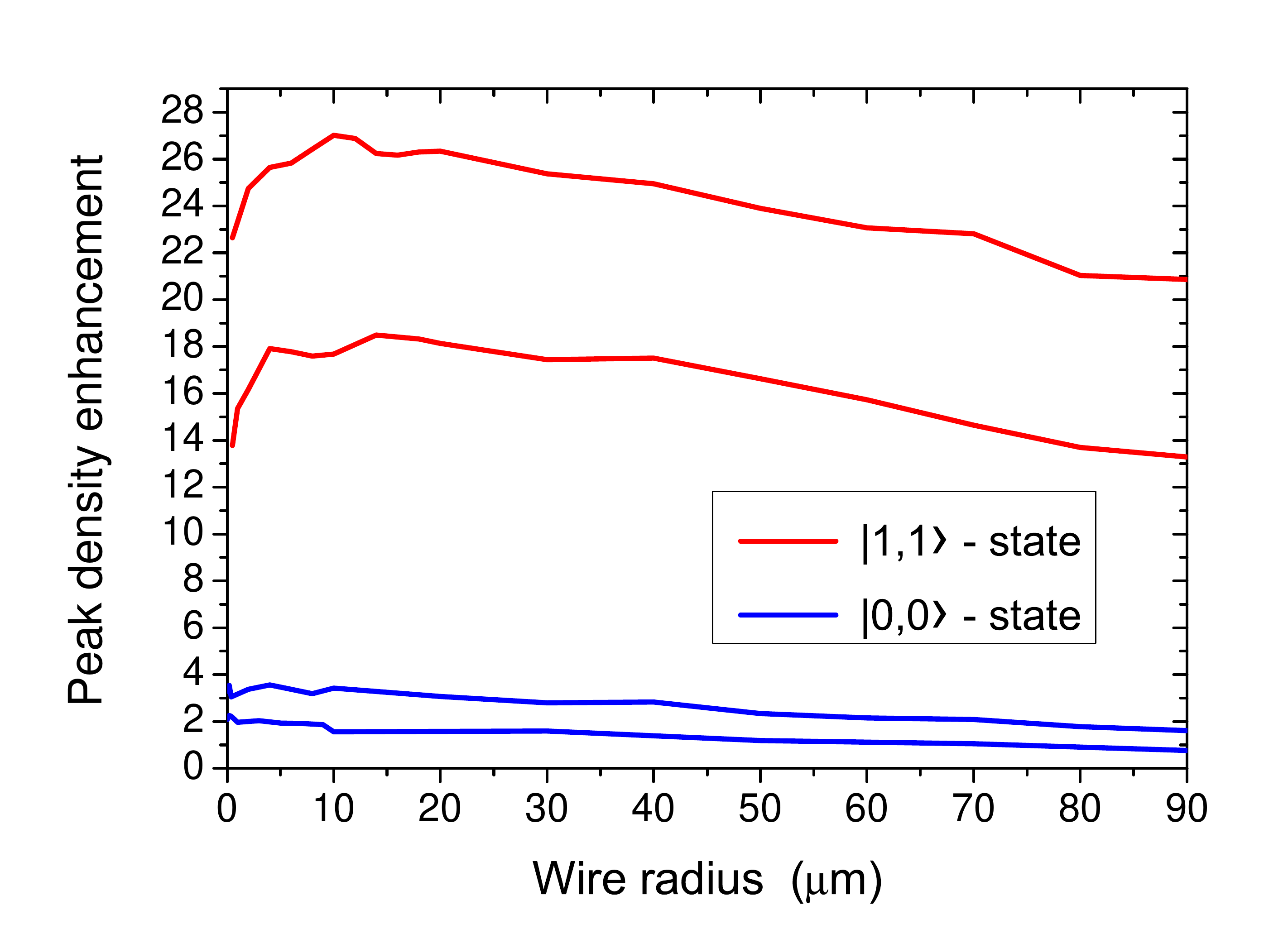}}
\caption{Simulated behavior of the ND$_3$ peak density due to guiding by the charged wire as a function of the wire radius for $\Delta x=375\,\mu$m and $\Delta x=475\,\mu$m (upper and lower lines, respectively). The beam velocity and the wire voltage are fixed to $v_0=310\,$m/s and $U=1.5\,$V, respectively.}
\label{fig:radius}
\end{center}
\end{figure}
Since the electric field which is generated by applying a certain voltage $U$ to the wire depends on the wire radius $R_w$ according to Eq.~\ref{eq:Efield}, the choice of $R_w$ crucially influences the guide efficiency.  In order to determine the optimum value of $R_w$ for the given conditions in our experiment we have performed simulations of the density enhancement for ND$_3$ at $U=1.5\,$kV and $v_0=310\,$m/s for various values of $R_w$ ranging from $90\,\mu$m down to about $0.5\,\mu$m (Fig.~\ref{fig:radius}). The two lines for either $|0,0\rangle$ and $|1,1\rangle$-states represent the results for $\Delta x=375\,\mu$m and $\Delta x=475\,\mu$m. As $R_w$ is reduced down to $\sim 10\,\mu$m the guiding efficiency first rises in spite of the decreasing electric field (Eq.~\ref{eq:Efield}) due to an increased number of trajectories that circle around the wire without hitting it. A clear maximum occurs around $10\,\mu$m corresponding to an optimum guiding efficiency for molecules in the $|1,1\rangle$-state. For smaller values of $R_w\lesssim10\,\mu$m, the guiding efficiency goes down again due to a sharp drop of the electric field. However, for practical reasons such as stability against tensile stress and electric sparkover we chose the BeCu wire with $50\,\mu$m diameter.

\begin{figure}
\begin{center}{
\includegraphics[width=0.49\textwidth]{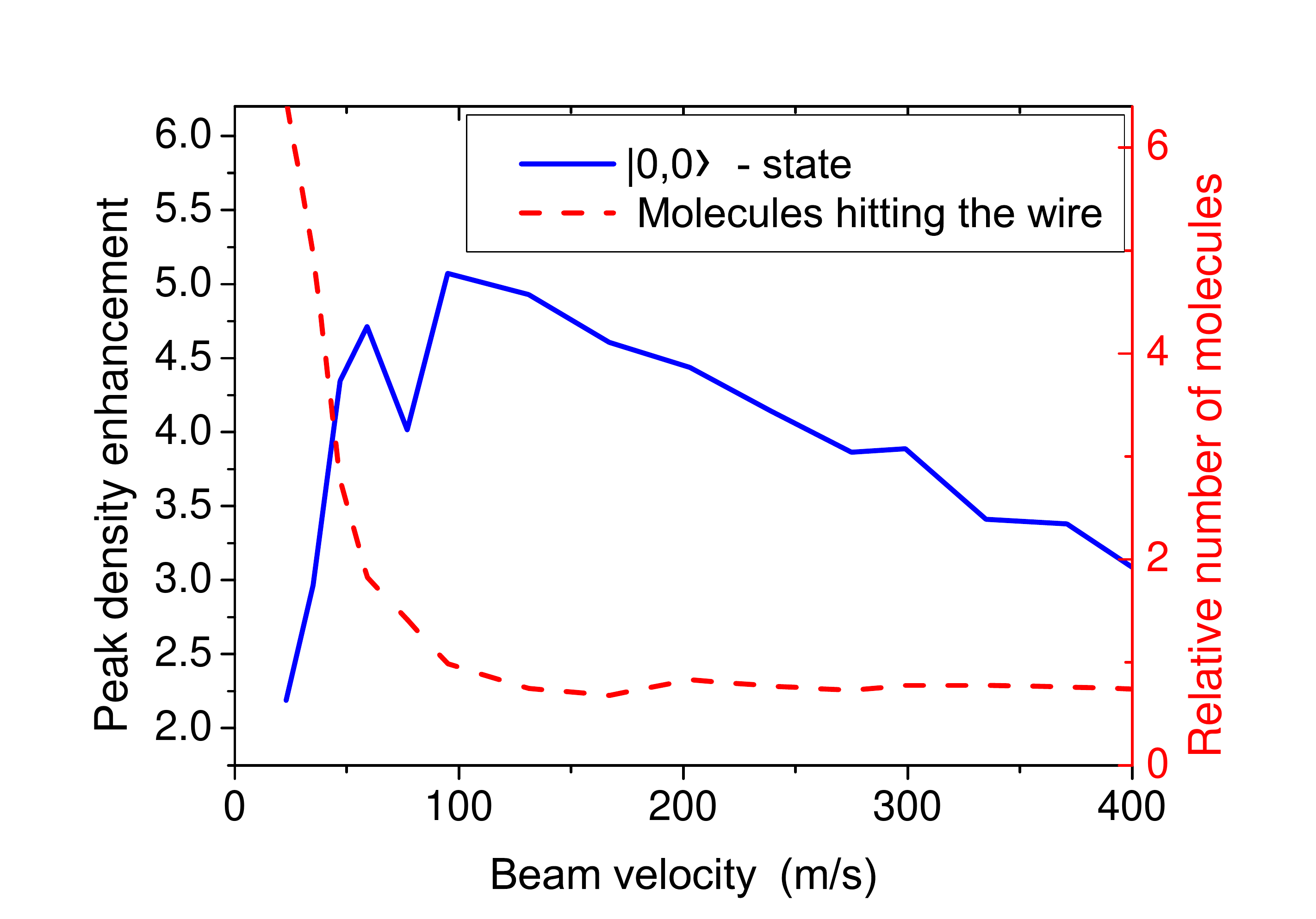}}
\caption{Simulated enhancement of the density of ND$_3$ molecules in the ground state $|0,0\rangle$ as a function of velocity down to $v_0\sim 20\,$m/s (solid line). The dashed line indicates the number of molecules that pitch into the wire on their way from the nozzle to the detector in proportion to the number of detected molecules.}
\label{fig:slow}
\end{center}
\end{figure}
The simulations presented so far show that ND$_3$ in both the linear $|1,1\rangle$ and the quadratic Stark state $|0,0\rangle$ can be guided along the charged wire in spite of the trajectories in a $r^{-2}$-potential being unstable. Therefore the question arises, under which conditions this instability becomes apparent as reduced guiding efficiency. In order to illustrate this effect we simulate the enhancement of the density of ND$_3$ molecules in the ground state $|0,0\rangle$ for beam velocities reaching down to $v_0\sim 20\,$m/s (Fig.~\ref{fig:slow}). The wire voltage $U=1.5\,$kV, the nozzle position $\Delta x=350\,\mu$m and the wire radius are held constant. As the velocity is decreased from $v_0=400\,$m/s down to $v_0=100\,$m/s the density enhancement first rises similarly to the state-averaged density shown in Fig.~\ref{fig:velocity}. However, at lower beam velocities $v_0<100\,$m/s the guiding effect sharply breaks down due to an increasing fraction of molecules that pitch into the wire (dashed line). Possibly the position of the maximum shifts down to lower velocities as the wire radius is further reduced. Thus we conclude that while the charged wire may be useful for guiding beams of ground state molecules at moderate velocities in the range of hundreds of m/s, is is inapplicable for molecules decelerated well below $v_0\sim100\,$m/s.

\section{Summary and outlook}
\label{sec:Conclusion}
In conclusion, we have demonstrated that a simple thin charged wire placed along the axis of a beam of decelerated polar $ND_3$ molecules can be used for enhancing the molecule density even at the exit of the wire guide by up to a factor 7. $ND_3$ and $CHF_3$, that was also used, show similar guiding behavior for comparable experimental parameters. Classical trajectory simulations are found to be in reasonable quantitative agreement with the measurements and to reproduce well the systematic trends. While rotational states with linear Stark effect are guided along the wire in stable Kepler orbits the quadratic ground state is only transiently enhanced at moderate beam velocities that ensue short flight times. Possible technical improvements of the present setup include a new nozzle design with a thinned end cap of the titanium ferrule in order to reduce the distance between the nozzle orifice and the wire surface down to $\sim100\,\mu$m. In this way both the guiding efficiency with respect to $U=0$ is enhanced and the absolute beam intensity is increased due to better beam alignment. Furthermore, the wire voltage could be increased up to $U\gtrsim3\,$kV by insulating the tip of the rotor against the vacuum apparatus to suppress sparking. In total, higher peak densities by about a factor $\gtrsim5$ as compared to the reported guided beam densities should be attainable.

Several applications of the presented concept are conceivable. Since only polar molecules experience the electrostatic guiding force in contrast to the rare-gas atoms in the carrier gas of the seeded expansion such a guiding element can be used to separate the molecules out of the carrier gas beam, in particular if an additional bend section is incorporated. The implementation of a wire guide into novel intense sources of cold molecules that rely on buffer gas cooling in combination with expansion cooling may be promising~\cite{Hsin:2011,Hutzler:2011}. Such an approach may be particularly advantageous in applications where decelerated cold molecules are to be injected into micro-structured devices such as micro-cavities or atom chips.

\begin{acknowledgments}
We thank H.\,J. Loesch for stimulation discussions. We are grateful for support by the Landesstiftung Baden-W\"urttemberg as well as by DFG.
\end{acknowledgments}

%\newpage %Just because of unusual number of tables stacked at end
%\bibliography{MarcelBib}% Produces the bibliography via BibTeX.

\end{document}